\documentclass[12pt]{article}

\usepackage[body={17.5cm, 22.5cm},right=2cm]{geometry}

\usepackage{float}

\usepackage{booktabs}
\usepackage{caption}

\usepackage{color}
\usepackage{graphicx}
\usepackage{epsf}
\usepackage{graphicx,epsfig}
\usepackage{amsmath}
\DeclareMathOperator{\arcsinh}{arcsinh}

\DeclareMathOperator{\arccosh}{arccosh}
 
\usepackage{multirow}

\usepackage{amsmath, amsthm, amssymb}

\usepackage{epsfig}
\usepackage{cite}
\usepackage{color,colordvi}

\newcounter{hran}

\makeatletter
\renewcommand\section{\@startsection {section}{1}{\z@}%
                               {-3.5ex \@plus -1ex \@minus -.2ex}%
                               {2.3ex \@plus.2ex}%
                               {\normalfont\large\bfseries}}
\makeatother

\begin{document}\thispagestyle{empty}

\vspace{0.5cm}

\def\thefootnote{\arabic{footnote}}
\setcounter{footnote}{0}

\def\s{\sigma}
\def\nn{\nonumber}
\def\p{\partial}
\def\ls{\left[}
\def\rs{\right]}
\def\lc{\left\{}
\def\rc{\right\}}

\newcommand{\be}{\begin{eqnarray}}
\newcommand{\ee}{\end{eqnarray}}
\newcommand{\bi}{\begin{itemize}}
\newcommand{\ei}{\end{itemize}}
\renewcommand{\th}{\theta}
\newcommand{\bth}{\overline{\theta}}

\vspace{0.5cm}
\begin{center}

{\large \bf 
On the Starobinsky Model of Inflation from  Supergravity
}
\\[1.5cm]
{\large  F. Farakos$^{a}$, A. Kehagias$^{a,b}$ and A. Riotto$^{b}$}
\\[0.5cm]

\vspace{.3cm}
{\normalsize {\it  $^{a}$ Physics Division, National Technical University of Athens, \\15780 Zografou Campus, Athens, Greece}}\\

\vspace{.3cm}
{\normalsize { \it $^{b}$ Department of Theoretical Physics and Center for Astroparticle Physics (CAP)\\ 24 quai E. Ansermet, CH-1211 Geneva 4, Switzerland}}\\

\vspace{.3cm}

\end{center}

\vspace{3cm}

\hrule \vspace{0.3cm}
{\small  \noindent \textbf{Abstract} \\[0.3cm]
\noindent 
We discuss how the higher-derivative Starobinsky model of inflation originates 
from  ${\cal{N}}=1$ supergravity. It is known that,  in the old-minimal supergravity
description written 
by employing a chiral compensator in the superconformal framework, the Starobinsky model  is equivalent to a no-scale model 
with $F$-term potential. We show that the Starobinsky model can also be originated 
within the so-called  new-minimal supergravity, where a linear compensator superfield 
is employed.
In this formulation, the  
Starobinsky model  is equivalent to standard supergravity coupled to a massive vector multiplet whose  
lowest scalar component plays the role of the inflaton and the vacuum energy is provided by  a  
 $D$-term potential. We also point out that higher-order  corrections to the  supergravity
 Lagrangian represent a threat to the   Starobinsky model as they can destroy the flatness of the inflaton potential in its scalar field
 equivalent description.

\vspace{0.5cm}  \hrule
\vskip 1cm

\def\thefootnote{\arabic{footnote}}
\setcounter{footnote}{0}



\baselineskip= 19pt
\newpage 
\section{Introduction}\pagenumbering{arabic}
If the perturbations during inflation \cite{lr} are originated by the same field driving inflation, 
the inflaton, then the recent Planck data on the cosmic microwave background radiation anisotropies 
have severely constrained the models of single-field inflation \cite{Ade:2013uln}. 
Indeed, successful models have to predict a significant red tilt in the two-point correlator of the scalar  
curvature perturbation, measured 
by the spectral index $n_s=0.960\pm 0.007$, and a low enough amount of tensor perturbations quantified by the 
current bound on the tensor-to-scalar ratio,   $r<0.08$. One of the models which better passes these constraints is 
the higher-derivative $R^2$ Starobinsky model \cite{star}. It is described by the Lagrangian 
(we set from now on the reduced Planckian mass to unity)
\be
\label{R^2} {\cal L}_{\text{star}} =\sqrt{-g}\left(R + \lambda_0  R^2\right), \,\,\, \ \lambda_0 > 0
\ee 
and it contains,  besides the graviton, one additional degree of freedom. 
The coupling constant $\lambda_0$ is positive in order to avoid instabilities.
Indeed, one can  rewrite the Lagrangian (\ref{R^2}) as \cite{Whitt:1984pd}
\be
\label{R2}
 {\cal L}_{\text{star}} =  \sqrt{-g}\left(R + \lambda_0  R\psi-\frac{1}{4} \lambda_0 \psi^2\right)
\ee
and, upon integrating out $\psi$, one gets back the original theory (\ref{R^2}).
Note that this is a classical equivalence. 
After writing the expression (\ref{R2}) in the Einstein
frame by means of the conformal transformation
\be
g_{\mu\nu}\to e^{-2\phi}g_{\mu\nu}=\left(1+\lambda_0\psi\right)^{-1} g_{\mu\nu}, 
\ee
we get the equivalent scalar field version of the Starobinsky model 
\be
 {\cal L}_{\text{star}} =  \sqrt{-g}\left[R-6\partial_\mu\phi\partial^\mu \phi-\frac{1}{4\lambda_0}(1-e^{-2\phi})^2\right]. \label{R3}
\ee
and the positivity of $\lambda_0$ is now obvious.
Inflation takes place when the scalar field is slowly-rolling along its potential plateau obtained for $\phi\gg 1$ and in order to achieve a sufficient number of e-folds the plateau must be at least as wide as ${\cal O }(5)$ in Planckian units.

In this paper we investigate the possibility of embedding   the  Starobinsky model into superconformal
theory and ${\cal{N}}=1$  supergravity.  
This extension is not unique. 
The reason is that there are two ways for 
the graviton, sitting along with the gravitino,  to fill a supergravity multiplet and one needs  a set of auxiliary fields
to define the off-shell supergravity multiplet. The minimal case should contain only the gravitino as fermionic content. This 
 means that we need a total of 12 bosonic degrees of freedom to match the 12 degrees of freedom of the gravitino. 
This is the so-called (12+12) supergravity theories and there are two of them\cite{Gates:1983nr,Buchbinder:1995uq}: the old-minimal supergravity and the new-minimal one.
The auxiliary fields of the old-minimal supergravity are a complex scalar and a vector, whereas the new-minimal supergravity
has a gauge one-form (gauging an $R$-symmetry) and a gauge two-form field. In particular, at the two derivative level, these 
two minimal supergravities are the same as they are related by some duality transformation of their auxiliary sectors. However,
when higher derivatives appear, this duality does not work and the two theories are different.  
Earlier \cite{Cecotti:1987sa,ketov} and recent \cite{Ellis,linde,buch} embeddings of the Starobinsky model have been all based on the old-minimal
formulation of ${\cal{N}}=1$ supergravity. 

An appropriate framework to discuss minimal supergravities is the superconformal calculus \cite{conf,Gates:1983nr,Buchbinder:1995uq,butter} 
which we employ here. 
To go to the desired Poincar\'e supergravity one fixes the appropriate compensator field and breaks the conformal symmetry.
This framework offers a connection between the different auxiliary field structure of the minimal Poincar\'e supergravities \cite{aux}.
Depending on the compensator, after gauge fixing
the superconformal symmetry, one recovers either old or new-minimal supergravity: with a chiral compensator, the old-minimal
supergravity is obtained whereas with  a real linear compensator superfield the new-minimal one is recovered.

The goal of this paper is two-fold: on one side, we wish to show that the Starobinsky model can be derived also from the
new-minimal formulation of supergravity in such a way that the vacuum energy driving inflation can be  identified with a $D$-term; on the other hand we want to point out that the  embedding of the Starobinsky model both in old- and new-minimal  supergravity
suffers of a potential problem deriving from the presence of higher-order corrections which may spoil the plateau of the potential of the scalar field driving inflation. This is reminiscent of the so-called $\eta$-problem \cite{lr} which arises when a model of inflation is embedded in  supersymmetry and the flatness of the potential is usually  spoiled by supergravity corrections \cite{lr}.

The paper is organized as follows. We describe the embedding  of the Starobinsky model within the old-minimal supergravity formulation in section 2 and within the new-minimal supergravity formulation in section 3. In section 4 we describe the potential danger represented by higher-order corrections in both formulations. Finally, we conclude in section 5.

\section{Starobinsky model  in the old-minimal supergravity }
We start by writing the Lagrangian that is appropriate to reproduce 
the supergravity version of the  Starobinsky model in the old-minimal framework
\be
\label{OM1}
{\cal L} = -3 [S_0 \overline{S}_0]_D +3 \lambda_1 [ {\cal R} {\cal \overline{R}} ]_D,
\ee
with 
\be
\label{R}
{\cal R} =-\frac{1}{4} S_0^{-1} \overline{\nabla}^2 \overline{S}_0.
\ee
Here  $S_0$ is the compensator chiral superfield, with scaling weight equal to $1$ and chiral weight $2/3$, 
the curvature chiral superfield ${\cal R}$ has scaling weight equal to $1$ and chiral weight $2/3$ as well, 
and $[{\cal{O}}]_D$ is the standard $D$-term density formula of conformal supergravity \cite{butter}, 
where ${\cal{O}}$ is a real superfield with scaling weight $2$ and vanishing chiral weight.
After gauge fixing the superconformal symmetry and choosing
\be
\label{S_0}
S_0 = 1, 
\ee
the superspace geometry is described by the old-minimal formulation, 
see for example Ref. \cite{W&B}.
Then Eq. (\ref{OM1}) becomes
\be
\label{OM1P}
{\cal{L}}= -3 \int {\rm d}^2 \Theta \, \, 2 \, {\cal{E}} \left\{ {\cal{R}} 
+\frac{\lambda_1}{8} \Big{(} \overline{{\cal{D}}}\overline{{\cal{D}}} - 8 {\cal{R}} \Big{)} \Big{(} {\cal{R}} \overline{{\cal{R}}}  \Big{)} 
\right\}+ {\rm h.c.}
\ee
It is easy to verify that the bosonic part 
of  Eq. (\ref{OM1P}) contains the Lagrangian (\ref{R^2})
\be
{\cal L}\supset R+\lambda_1 R^2
\ee
and therefore is a good candidate for the supergravity theory we are after \cite{R2D2}. 
The next step is to write the expression (\ref{OM1}) as  standard supergravity with additional degrees of freedom
in the same way we have traded the $R^2$ term in non-supersymmetric case (\ref{R^2}) by a scalar field 
coupled to Einstein gravity in (\ref{R2}). This can be implemented by using appropriate Lagrange multipliers. 
Hence,
we introduce a chiral superfield $J$ with scaling weight $1$ (chiral weight 2/3) 
and a chiral Lagrange multiplier $\Lambda$ with scaling weight $2$ (chiral weight 4/3) 
and the equivalent Lagrangian to (\ref{OM1}) is
 \cite{Cecotti:1987sa}
\be
\label{OM2}
{\cal L} = -3 [S_0 \overline{S}_0]_D +3 \lambda_1 [ J \overline{J} ]_D +3 ( [ \Lambda ( J - {\cal R}  )]_F +{\rm h.c.} )
\ee
where $[{\cal{O}}]_F$ is the standard $F$-term density formula of conformal supergravity \cite{butter}, 
with ${\cal{O}}$ a chiral superfield having scaling weight $3$ and chiral weight $2$.
Indeed, integrating out the Lagrange multiplier chiral superfield $\Lambda$ from Eq. (\ref{OM2}) we get
\be
\label{J1}
J = {\cal R} 
\ee
and  Eq.  (\ref{OM1}) is reproduced by Eq. (\ref{OM2}). 
By using the identity 
\be
\label{FtoD}
[ \Lambda  {\cal R}    ]_F =  [ \Lambda \overline{S_0} S_0^{-1} ]_D, 
\ee
Eq. (\ref{OM2}) can be recast in the form 
\be
\label{OM3}
{\cal L} = -3 [S_0 \overline{S}_0]_D +3 \lambda_1 [ J  \overline{J} ]_D 
-3 [ \Lambda \overline{S_0} S_0^{-1} ]_D - 3[ \overline{\Lambda} S_0 \overline{S}_0^{-1} ]_D
+3 ( [\Lambda J ]_F +{\rm h.c.} ),
\ee
and will lead to standard Poincar\'e supergravity.
By defining new chiral superfields ${\cal C}$ and ${\cal T}$ defined in terms of our original $J$ and $\Lambda$ as 
\be
\label{TC}
{\cal C}= \sqrt{\lambda_1} S_0^{-1} J  \ ,~~~ \  {\cal T} = \frac{1}{2} + S_0^{-2} \Lambda,
\ee
Eq. (\ref{OM3}) turns out to be
\be
\label{OM4}
{\cal L} = -3 \Big{[}S_0 \overline{S}_0 ( {\cal T} + \overline{{\cal T}} - {\cal C} \overline{{\cal C}} )\Big{]}_D 
+ 3\lambda_1^{-1/2} \left( \left[ {\cal C} \, \left({\cal T} -\frac{1}{2} \right) S_0^3 \right]_F +{\rm h.c.}\right).
\ee
We recognize in  Eq. (\ref{OM4}) the characteristic form of a no-scale model \cite{Cecotti:1987sa}. In particular, the fields
${\cal C}$ and ${\cal T}$ parametrize the scalar manifold $SU(2,1)/U(2)$. 
Note that the theory  is not gauged and  the potential is due to an $F$-term, the second term in (\ref{OM4}).
We now gauge fix the superconformal symmetry in order for the 
superspace to be described by the old-minimal formulation.
Then Eq. (\ref{OM4}) turns out to be  
the standard old-minimal supergravity Lagrangian coupled to chiral superfields
\be
\label{OM5}
{\cal{L}}=\int {\rm d}^2 \Theta \, \, 2 \, {\cal{E}} \left\{
\frac{3}{8} \Big{(} \overline{{\cal{D}}}\overline{{\cal{D}}} - 8 {\cal{R}} \Big{)} e^{ - K/3 } + W
\right\}+ {\rm h.c.}
\ee
with K\"ahler potential
\be
\label{K1}
K = -3\, \ln\Big{(}{\cal T} + \overline{{\cal T}} - {\cal C} \overline{{\cal C}}\Big{)} 
\ee
and superpotential
\be
\label{W1}
W=  \frac{3}{\sqrt{\lambda_1}} {\cal C} \left({\cal T} - \frac{1}{2}\right).
\ee
The bosonic sector of the final Lagrangian is
\be
\label{OM6}
e^{-1}{\cal L} = \frac{1}{2} R  - K_{i {\overline{j}}} \partial_{\mu} z^i \partial^{\mu} \overline{z}^{\overline{j}} 
-\ V_F
\ee
with 
\be
\label{VF1}
V_F =  e^{K} \left[ K^{i {\overline{j}}} (D_iW) (D_{\overline{j}}\overline{W}) -3 W \overline{W} \right], ~~~ \, i,j=1,2, 
\ee
where $z^1=T$ and $z^2=C$ the scalar lowest components of the chiral superfields ${\cal T}$ and ${\cal C}$.
We have used the standard notation 
\be 
K_i=\frac{\partial K }{ \partial z^i}\, , ~~~K_{i {\overline{j}}} = 
\frac{\partial^2 K(z,\overline{z})}{\partial z^i \partial\overline{z}^{\overline{j}}}\, , ~~~~
 \label{DW} D_iW=W_i +K_iW ,~~~~W_i=\frac{\partial W }{ \partial z^i}.
 \ee 

The superpotential (\ref{W1}) belongs to a specific class of supergravity theories 
studied in \cite{kal-lin}, where together with 
a K\"ahler potential invariant under $C\to - C$,  a local extremum at $C=0$ appears. 
This also happens in our case as there 
is a local extremum at $C={\rm Im}\,T=0$ where we have the inflationary potential. 
Indeed, by parametrizing  the complex scalar $T$ by two real scalar $\phi,b$, as
\be
\label{T}
T=\frac{1}{2}e^{\frac{2}{\sqrt{3}} \phi} + i b, 
\ee
we find that there is an extremum at 
$C=b =0,$
where the effective bosonic theory turns  out to be 
\be
\label{OM7}
e^{-1}{\cal L} = \frac{1}{2} R  -   \partial_{\mu} \phi  \partial^{\mu} \phi 
-\ \frac{3}{2 \lambda_1 } \,  \left(1-e^{-\frac{2}{\sqrt{3}} \phi}\right)^2 .
\ee
This is just the Starobinsky theory formulated in terms of the extra scalar degree of freedom. 
However, there is a possibility  of a tachyonic instability for 
excitations along 
the inflationary trajectory $C=0$. Indeed, the mass of such excitations are \cite{kal-lin}
\be
m^2_{\pm}=-\left(K_{CC\bar{C}\bar{C}}\pm \big{|}K_{CC\bar{C}\bar{C}}-K_{CC}\big{|}\right)|f|^2+|\partial_Tf|^2
\ee
for a general superpotential $W=C f(T)$ and a K\"ahler potential invariant under $C\to -C$. It is easy to check that 
in our case we have in fact a tachyonic instability during the inflationary phase. A remedy can be  modifying 
the K\"ahler potential appropriately \cite{linde}. We may consider, for example, instead of (\ref{OM1}) the theory
\be
\label{OM21}
{\cal L} = -3 [S_0 \overline{S}_0]_D +3 \lambda_1 [ {\cal R} {\cal \overline{R}} ]_D 
+ 3 \zeta \, [{\cal R} {\cal \overline{R}} \ {\cal F}\left({\cal R} {\cal \overline{R}} (S_0\overline{S}_0)^{-1} \right) ]_D .
\ee
After writing this theory as   standard supergravity as we have done for (\ref{OM1}) and gauge fixing $S_0 =1$, 
the new term does not change the superpotential and  changes only the corresponding K\"ahler potential,
which turns out  to be 
\be
\label{K7}
K = -3\, \text{ln}\,\big{(} {\cal T} + \overline{{\cal T}} - {\cal C} \overline{{\cal C}} [ 1 
+ \zeta \, \lambda_1^{\, -1} \,  {\cal F} ({\cal C} \overline{{\cal C}} \, \lambda_1^{\, -1}) ] \big{) }. 
\ee
As suggested in Ref. \cite{linde}, the choice  
\be
{\cal F} = - \lambda_1 {\cal C} \overline{{\cal C}}+\cdots 
\ee
will stabilize the inflationary trajectory and give rise to a consistent theory for appropriate values of $\zeta$. 
A different approach to Starobinsky inflation in supergravity has been
followed in  Refs. \cite{ketov} where  appropriate F-supergravity
extension of f(R) gravity has been employed, dual to the standard approach
in terms of
K\"ahler potentials and superpotentials.

Let us now turn to the alternative derivation of the same Lagrangian in the new-minimal supergravity formulation.

\section{Starobinsky model in new-minimal supergravity}
In this section we want to show that there  is another way to write a supergravity which contains the Lagrangian (\ref{R^2}) in its bosonic sector. 
The appropriate compensator for the new-minimal supergravity gauging 
is a real linear multiplet ($L_0$) with scaling weight $2$ and vanishing weight under chiral rotations.
We employ now the following Lagrangian 
\be
\label{NM1}
{\cal L} = [ L_0 V_R ]_D +  \frac{\lambda_2}{4} ( [ W^{\alpha}(V_R) W_{\alpha}(V_R)]_F +{\rm h.c.} ), 
\ee
where 
\be
\label{VR}
V_R &=& \text{ln}\,\left(\frac{L_0}{Y_0\overline{Y}_0}\right),
\\
\label{VRW}
W_{\alpha}(V_R) &=&-\frac{1}{4} \overline{\nabla}^2 \, \nabla_{\alpha} (  V_R ) 
\ee
and $Y_0$ a chiral superfield with scaling weight $1$. 
After gauge fixing the superconformal symmetry and choosing
\be
\label{L_0}
L_0 = 1 
\ee
the superspace geometry is described by the new-minimal formulation, 
see for example Refs. \cite{Sohnius:1982fw,Ferrara:1988qxa}.
Indeed, fixing the superconformal symmetry by  $L_0=1$, we get that the
 graviton multiplet contains  four fields, the physical graviton $e_\mu^a$,
the gravitino $\psi_\mu$ and two auxiliary gauge fields $A_\mu$,
and $B_{\mu\nu}$ with corresponding gauge invariances
\be
\delta A_\mu=\partial_\mu\phi\, , ~~~~\delta b_{\mu\nu}=\partial_\mu b_\nu-\partial_\nu b_\mu.
\ee
In fact, $A_\mu$ gauges the $U(1)_R$ $R$-symmetry of the superconformal algebra, which survives 
after the gauge fixing (\ref{L_0}). 
Then, the desired theory is described by the following new-minimal Poincar\'e superspace density
\be
\label{NM1f}
{\cal{L}}=\int {\rm d}^2 \Theta  \, \, 2 \, {\cal{E}} \left\{
- \frac{1}{8}  \overline{{\cal{D}}}\overline{{\cal{D}}} V_R 
+ \frac{\lambda_2}{4} \, W^{\alpha}(V_R) W_{\alpha}(V_R)  \right\}+ {\rm h.c.}, 
\ee
where now $V_R$ turns out to be the gauge multiplet of the supersymmetry algebra, namely
\begin{eqnarray}
\label{gaugemult}
V_R=\left(- H_\mu + \frac{1}{3} A_\mu,
- \frac{1}{3}\gamma_5\gamma^\nu 
r_\nu,
-\frac{1}{6}\hat{\cal{R}} - H_\mu H^\mu \right),
\end{eqnarray}
where $r_\nu$ is the supercovariant gravitino field strength, $\hat{R}$ is the (supercovariant) Ricci scalar
and $H^\mu$ the Hodge dual of the (supercovariant) field strength for the auxiliary two-form \cite{Ferrara:1988qxa}. 
The first terms in Eq. (\ref{NM1f}) is easily recognized as the Fayet-Iliopoulos term for the gauge multiplet,
whereas the second is its standard kinetic term. 
Since the highest component $D_R$ of the gauge multiplet contains the Ricci
scalar ($D_R\sim R$), clearly we will get the desired 
$D_R+\lambda_2 D_R^2\sim R+\lambda_2 R^2$ from the terms in  Eq. (\ref{NM1f}).
See Ref. \cite{Cecotti:1987qe} for a thorough discussion.

As a first step to write Eq. (\ref{NM1}) as standard Poincar\'e supergravity, 
we consider $L_0$ as an unconstrained real superfield 
(note that by employing the equation of motion for $Y_0$ we can make $L_0$ real linear again).
Then one can check that the following Lagrangian 
\be
\label{NM2}
{\cal L} = [ L_0 V_R ]_D + \frac{\lambda_2}{4} ( [ W^{\alpha}(V) W_{\alpha}(V)]_F +{\rm h.c.} ) + [L' ( V - V_R)]_D
\ee
reproduces Eq. (\ref{NM1}) when we integrate out the real linear superfield $L'$ to find
\be
\label{L'1}
V= \text{ln}\,\left(\frac{L_0}{Y_0\overline{Y}_0}\right) - \text{ln}\,\Phi  - \text{ln}\,\overline{\Phi}  +c
\ee
and plug it back into Eq. (\ref{NM2}).
Now, in order to write the theory as standard supergravity, we go in the opposite direction.
We again perform a variation with respect to $L'$, but now we interpret the equation of motion as
\be
\label{L'2}
\text{ln}\,\left(\frac{L_0}{Y_0\overline{Y}_0}\right) = V + \text{ln}\,\Phi  + \text{ln}\,\overline{\Phi}  + c,
\ee
which can be solved for $L_0$ by
\be
\label{L'3}
\frac{L_0}{Y_0\overline{Y}_0} = \overline{\Phi} e^{V + c} \Phi.
\ee
The final step is to plug back Eq. (\ref{L'3}) (or (\ref{L'2})) into Eq. (\ref{NM2}) to get 
\be
\label{NM3}
{\cal L} = [ Y_0 \overline{Y}_0 ( \overline{\Phi} e^{V + c} \Phi \ \text{ln}\,( \overline{\Phi} e^{V + c} \Phi ) )]_D 
+\frac{\lambda_2}{4} ( [ W^{\alpha}(V) W_{\alpha}(V)]_F +{\rm h.c.} ).
\ee
The action (\ref{NM3}) is the dual action to (\ref{NM1}) 
\cite{Cecotti:1987qr,Cecotti:1987qe}.
Since $c$ is just an integration constant we can take $c=0$.
Note that our theory here is gauged and that the potential is thus due to the standard $D$-term, in contrast to the expression (\ref{OM4}).
Again gauge fixing superconformal invariance and setting 
$$Y_0 = 1,$$ 
we recover the following standard ${\cal{N}}=1$ supergravity 
\be
\label{NM4}
{\cal{L}}=\int {\rm d}^2 \Theta \, \, 2 \, {\cal{E}} \left\{
\frac{3}{8} \Big{(} \overline{{\cal{D}}}\overline{{\cal{D}}} - 8 {\cal{R}} \Big{)} e^{ - K/3 } 
+ \frac{\lambda_2}{4} \, W^{\alpha} W_{\alpha}  \right\}+ {\rm h.c.}, 
\ee
with the K\"ahler potential
\be
\label{K2}
K = -3 \ln \left[-\frac{1}{3} \overline{\Phi} e^{V} \Phi \ln( \overline{\Phi} e^{V} \Phi )  \right]. 
\ee
In component form the expression (\ref{NM4}) reads (after rescaling $V \rightarrow 2V$)
\be
\nn
e^{-1}{\cal{L}}&=& \frac{1}{2} R 
- K_{A {\overline{A}}} D_\mu A  \overline{D^\mu A} +\frac{1}{2} \left(K_A A + K_{\overline{A}} \overline{A}   \right)\, D
\\
\label{NM5}
 &-&2 \lambda_2  
\left(  \frac{1}{2} F^{\mu\nu} F_{\mu\nu}
-\frac{i}{4} \epsilon^{\mu\nu\rho\sigma} F_{\mu\nu} F_{\rho\sigma} 
- D^2 \right)
\ee
with 
\be
D_\mu A = \partial_\mu A +i A_\mu A,
\ee
and after  integrating out the auxiliary $D$ we get
\begin{align}
\label{NM6}
e^{-1}{\cal{L}}&=\frac{1}{2} R -2 \lambda_2  
\left(  \frac{1}{2} F^{\mu\nu} F_{\mu\nu}-
\frac{i}{4} \epsilon^{\mu\nu\rho\sigma} F_{\mu\nu} F_{\rho\sigma} 
 \right)
- \frac{3}{ A\overline{A}  \left[\ln (A\overline{A}) \right]^{2} } 
D_\mu A \,\overline{D_\mu A} 
-\frac{9}{8 \lambda_2} \left[ 1 + \frac{1}{\ln(A \overline{A} )}  \right]^2.
\end{align}
With the redefinition
\be
\ln A=  - \frac{1}{2} e^{\frac{2}{\sqrt{3}} \phi } + i a,
\ee
the expression (\ref{NM6}) is finally written as (with $\lambda_2=1/4g^2$)

\begin{align}
\label{NM7}
 e^{-1}{\cal{L}}& = \frac{1}{2} R - 
 \frac{1}{4g^2} F^{\mu\nu} F_{\mu\nu}+
\frac{i}{8g^2} \epsilon^{\mu\nu\rho\sigma} F_{\mu\nu} F_{\rho\sigma} 
- 3 e^{ - \frac{4}{\sqrt{3}} \phi }\big{(}\partial_\mu a + A_\mu\big{)}^2\nonumber \\
 &
 -\partial_\mu \phi \partial^\mu \phi
-\frac{9}{2 } g^2\left( 1 -   e^{-\frac{2}{\sqrt{3}} \phi }  \right)^2.
\end{align}
This describes a massive vector with mass  
\be
m_A=\sqrt{6} g\, e^{-2\phi/\sqrt{3}}
\ee
in Planck units, and a singlet scalar $\phi$. The latter can be considered as the inflaton field with a $D$-term potential
\be
\label{VD1}
 V_D = \frac{9}{2 }g^2 \left( 1 - e^{-\frac{2}{\sqrt{3}} \phi }  \right)^2. 
\ee
Therefore, the $R^2$ new-minimal supergravity is described by standard supergravity
coupled to a massive vector superfield. The latter contains a real scalar in its lowest component (the $\phi$ field here)
and a massive $U(1)$ vector in its bosonic sector. Thus, the Starobinsky model stems from  the new-minimal supergravity  constructed 
by means of a  massless vector multiplet and a chiral multiplet. The vector eats one of the scalars of the 
chiral multiplet and becomes massive, whereas the other scalar of the chiral acquires a $D$-term potential. All together,
the massless vector and the two scalars of the chiral, rearrange themselves such that to form standard supergravity coupled 
to a massive vector multiplet. Note that the scalar $\phi$ is what is usually fixed to zero by 
imposing the Wess-Zumino gauge in exact gauge invariance.

\section{The  issue  of higher-order corrections}
Before concluding, we  would like to discuss a relevant issue that might represent a potential danger to the embedding of the the
Starobinsky model into supergravity: higher order corrections. As we shall see, both in the old- and new-minimal supergravity formulation of the Starobinsky model, one can add non-renormalizable higher-order corrections  which are admitted by the symmetries and might spoil the plateau of the inflaton potential necessary to drive inflation.

\subsection{Corrections in new-minimal supergravity}
Let us first discuss the possible corrections to the inflaton potential (\ref{VD1}) obtained in the new-minimal version. 
These corrections are generated as  corrections
to the superconformal action (\ref{NM1}). However, 
all possible non-renormalizable terms are  restricted by
gauge invariance. Possible corrections could arise from higher-order $D$-terms of the supersymmetric field strength $W_\alpha$.
In conformal superspace we may consider 
\be
\nn
{\cal L} &=& [ L_0 V_R ]_D +  \frac{\lambda_2}{4} ( [ W^{\alpha}(V_R) W_{\alpha}(V_R)]_F +{\rm h.c.}  ) 
\\
\label{NM11}
&+& \frac{\xi}{16} [( W^{\alpha}(V_R) W_{\alpha}(V_R)  \overline{W}_{\dot{\alpha}}(V_R) \overline{W}^{\dot{\alpha}}(V_R)) (L_0)^{-2}]_D.
\ee
In the $L_0 = 1$ gauge, this theory will contain in it bosonic sector terms of the form
\be
\label{R4}
{\cal L } \supset R + \lambda_2 R^2 +\xi R^4, 
\ee
which represent corrections to Starobinsky theory 
 in the new-minimal supergravity framework.
To recover the dual theory, we write Eq. (\ref{NM11}) as
\be
\nn
{\cal L} &=& [ L_0 V_R ]_D +\frac{\lambda_2}{4} ( [ W^{\alpha}(V) W_{\alpha}(V)]_F +{\rm h.c.} ) 
\\
\label{NM12}
&+& \frac{\xi}{16} [( W^{\alpha}(V) W_{\alpha}(V)  \overline{W}_{\dot{\alpha}}(V) \overline{W}^{\dot{\alpha}}(V)) (L_0)^{-2}]_D+ [L' ( V - V_R)]_D.
\ee
Again we perform a variation with respect to $L'$, and we interpret the equation of motion as 
\be
\label{L''2}
\text{ln}\,\left(\frac{L_0}{Y_0\overline{Y}_0}\right) = V + \text{ln}\,\Phi  + \text{ln}\,\overline{\Phi}  +c,
\ee
which can be solved for $L_0$ 
\be
\label{L''3}
\frac{L_0}{Y_0\overline{Y}_0} = \overline{\Phi} e^{V } \Phi.
\ee
We have also   set $c=0$ here.
The final step is to plug back (\ref{L''3}) (or (\ref{L''2})) into (\ref{NM12}), which gives 
\be
\nn
{\cal L} &=& [ Y_0 \overline{Y}_0 ( \overline{\Phi} e^{V } \Phi \ \text{ln}\,( \overline{\Phi} e^{V } \Phi ) )]_D 
+\frac{\lambda_2}{4}( [ W^{\alpha}(V) W_{\alpha}(V)]_F +{\rm h.c.} )
\\
\label{NM13}
&+& \frac{\xi}{16} [( W^{\alpha}(V) W_{\alpha}(V)  \overline{W}_{\dot{\alpha}}(V) \overline{W}^{\dot{\alpha}}(V)) 
(Y_0 \overline{Y}_0)^{-2} (\overline{\Phi} e^{V } \Phi)^{-2}]_D.
\ee
We now gauge fix the conformal symmetry and set $Y_0 = 1$ to recover the standard  supergravity theory that corresponds to (\ref{NM13})
\begin{eqnarray}
{\cal{L}}&=&\int {\rm d}^2 \Theta  \, \, 2 \, {\cal{E}} \left\{
\frac{3}{8} \Big{(} \overline{{\cal{D}}}\overline{{\cal{D}}} - 8 {\cal{R}} \Big{)} e^{ - K/3 } 
+ \frac{\lambda_2}{4} \, W^{\alpha} W_{\alpha} \right.  \nonumber 
\\
&-&\left.\frac{1}{4}\Big{(}\overline{{\cal{D}}}\overline{{\cal{D}}} - 8 {\cal{R}}\Big{)} 
\left[ \frac{\xi}{32 (\overline{\Phi} e^{V }  \Phi)^2}  W^{\alpha} W_{\alpha}  \overline{W}_{\dot{\alpha}} \overline{W}^{\dot{\alpha}}\right] 
 \right\}+ {\rm h.c.},  \label{EP1}
\end{eqnarray}
with 
\be
\label{KEP1}
K &=& -3\,\text{ln}\,\left[-\frac{1}{3} \overline{\Phi} e^{V } \Phi \ \text{ln}\,( \overline{\Phi} e^{V } \Phi )\right].
\ee
Importantly, the K\"ahler potential is the same as in Eq. (\ref{K2}) and thus  there  are no corrections to 
the K\"ahler potential.  
In addition, the theory (\ref{EP1}) has been studied in Refs. \cite{Farakos:2013bca,Cecotti:1986jy}, 
where  now the general functions in the higher derivative gauge sector are fixed by the form of the integrated out $L_0$.
The component form reads (after rescaling $V \rightarrow 2V$)
\be
\nonumber
e^{-1}{\cal{L}}&=& \frac{1}{2} R 
- K_{A {\overline{A}}} D_\mu A  \overline{ D^\mu A} +\frac{1}{2} \left(K_A A + K_{\overline{A}} \overline{A}   \right) D
\\
\nn
 &-& 2 \lambda_2  
\left( \frac{1}{2} F^{\mu\nu} F_{\mu\nu}
- \frac{i}{4} \epsilon^{\mu\nu\rho\sigma} F_{\mu\nu} F_{\rho\sigma} 
- D^2 \right)
\\
\label{U1}
&+&\frac{\xi}{ (\overline{A}  A)^2} e^{ \frac{-2 K}{3}} 
\left[\frac{1}{4} (F^{\mu\nu} F_{\mu\nu})^2 - F^{\mu\nu} F_{\mu\nu} D^2 + 
\frac{1}{16} (\epsilon^{\mu\nu\rho\sigma} F_{\mu\nu} F_{\rho\sigma} )^2 +D^4 \right].
\ee
To find the scalar potential we have to integrate over $D$.
Since we are interested only in the scalar potential in what follows we 
ignore all other contributions to $D$, but those from $A$. For a more complete discussion, one may consult 
\cite{Farakos:2013bca,Cecotti:1986jy}.
By defining  the functions
\be
\label{a}
a&=& -\frac{1}{2} [K_A A + K_{\overline{A}} \overline{A}   ],
\\
\label{b}
b&=& 2 \lambda_2,
\\
\label{c}
c&=& \frac{\xi e^{ \frac{-2 K}{3}} }{ (\overline{A}  A)^2}, 
\ee
the equation of motion for $D$ turns out to be
\be
\label{eqD}
0 = a + 2 b D + 4 c D^3. 
\ee
The solution to Eq. (\ref{eqD}) was found in Ref.  \cite{Cecotti:1986jy}
and  is given by
\be
\label{solD}
D= \sqrt{\frac{2 b}{3 c}}  \sinh n,
\ee
with
\be
\label{n}
n = \frac{1}{3} \arcsinh \left(- \frac{3a}{4b} \sqrt{\frac{6 c}{b} }   \right).
\ee
The scalar potential reads
\be
\label{VD}
V_D=   \frac{4 \lambda_2 b}{3 c}  \cosh(2n) (\sinh n)^2.
\ee
\begin{figure}[H]
\begin{center}
\includegraphics[scale=0.6]{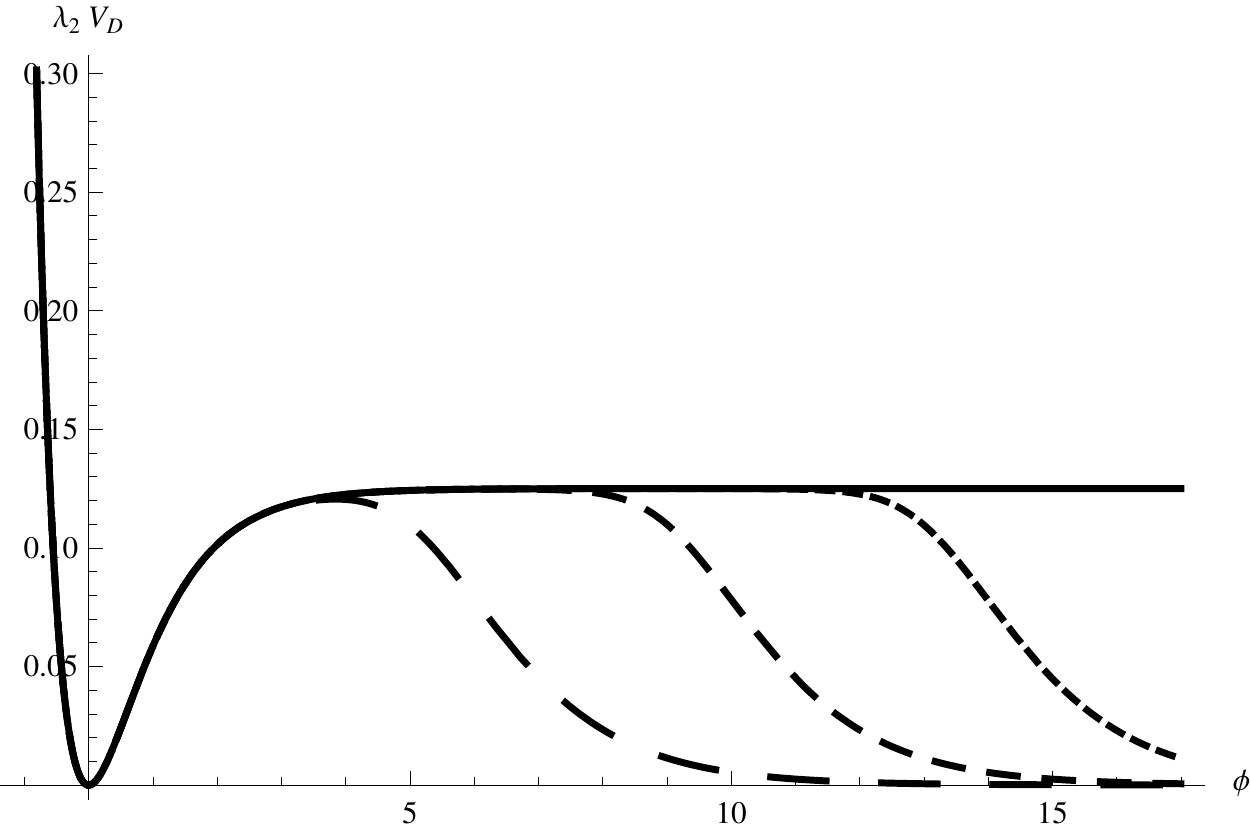}
\caption{\small The scalar potential   for three different values of $s=\frac{\sqrt{\xi}}{4}
\left(\frac{3}{\lambda_2}\right)^{3/2} $: \, $i$) $s= 10^{-2}$ (long dashed line), $ii$) $s=10^{-4}$ (medium dashed line) and $iii$)
$s=10^{-6}$ (short dashed line). The horizontal line corresponds to $\xi=0$.}
\end{center}
\end{figure}
\noindent
To find the corrections to the inflaton potential (\ref{VD1}), we rewrite $n$ as
\be
\label{nn}
n = \frac{1}{3} \arcsinh \omega,  
\ee
where
\be
\label{omega1}
\omega = \frac{\sqrt{\xi}}{8} \left(\frac{3}{\lambda_2}\right)^{\frac{3}{2}} \left(1 - e^{\frac{2}{\sqrt{3}} \phi}\right).
\ee
For  $\omega\ll  1$, the potential is written as 
\be
V_D\approx  \frac{9}{8 \lambda_2} \left( 1 - e^{-\frac{2}{\sqrt{3}} \phi }  \right)^2-\frac{9\xi}{256\lambda_2^4}
e^{-\frac{4}{\sqrt{3}} \phi } \left( 1 - e^{\frac{2}{\sqrt{3}} \phi }  \right)^4,  ~~~~~\omega\ll1.
\ee 
On the other side, if $|\omega|\gg1$ ($\phi\gg1$), the potential is approximately given by
\be
V_D\approx \frac{\lambda_2^2}{3\xi}e^{-\frac{4}{3\sqrt{3}} \phi },  ~~~~~\phi\gg 1. 
\ee
We have plotted the potential  in Fig. 1 for various values of the parameter
$s=\frac{\sqrt{\xi}}{4}
\left(\frac{3}{\lambda_2}\right)^{3/2} $. If $s=0$ (i.e., $\xi=0$), the potential has a plateau for large positive  values 
of $\phi$ and one recovers the nice feature of the Starobinsky model formulated in terms of the extra scalar degree of freedom. However, for  non-zero values of $s$, the plateau is restricted to smaller regions and it disappears for larger values of $s$ with a fall-off 
$V_D\sim e^{-\frac{4}{3\sqrt{3}} \phi }$ after the plateau. Therefore, the higher-order corrections pose a problem to the Starobinsky model: 
we know that successful inflation is achieved when the number of e-folds is about 60. This requires the field plateau to be as large as ${\cal O}(5)$ in Planckian units. This imposes the parameter $s$ to be smaller than about $10^{-4}$. Even so, one should explain why the initial value field is positioned on the plateau, instead of being on the fall-off region.


\subsection{Corrections in old-minimal supergravity}
Higher-order corrections are also expected in the old-minimal supergravity case.
It is straightforward to verify that the following superspace Lagrangian
\be
\label{OM11}
{\cal L} = -3 [S_0 \overline{S}_0]_D +3 \lambda_1 [ {\cal R} {\cal \overline{R}} ]_D 
+ \xi  [ \nabla^{\alpha} ({\cal R} / S_0) \nabla_{\alpha} ({\cal R} / S_0) 
\overline{\nabla}_{\dot{\alpha}}  ({\cal \overline{R}} / \overline{S}_0) \overline{\nabla}^{\dot{\alpha}}  ({\cal \overline{R}} / \overline{S}_0) ]_D 
\ee
reproduces (\ref{R4}) after gauge fixing $S_0=1$.
We can rewrite (\ref{OM11}) as 
\be
\label{OM12}
{\cal L} = -3 [S_0 \overline{S}_0]_D +3 \lambda_1 [ J  \overline{J} ]_D 
+ \xi  [ \nabla^{\alpha} (J / S_0) \nabla_{\alpha} (J / S_0) 
\overline{\nabla}_{\dot{\alpha}}  (\overline{J} / \overline{S}_0) \overline{\nabla}^{\dot{\alpha}} (\overline{J} / \overline{S}_0) ]_D 
+3( [ \Lambda ( J - {\cal R}  )]_F +{\rm h.c.} ).
\ee
Now,  making the redefinitions (\ref{TC}), the  theory (\ref{OM12}) becomes
\be
\nn
{\cal L} &=& -3 [S_0 \overline{S}_0 ( {\cal T} + \overline{{\cal T}} - {\cal C} \overline{{\cal C}} )]_D 
+ 3(\sqrt{\lambda_1})^{-1} ( [ {\cal C} \, ({\cal T} -\frac{1}{2} ) S_0^3 ]_F +{\rm h.c.})
\\
\label{OM13}
&+& \frac{\xi}{ \lambda_1^2} [ \nabla^{\alpha}  {\cal C} \nabla_{\alpha} {\cal C} 
\overline{\nabla}_{\dot{\alpha}}  \overline{{\cal C}} 
\overline{\nabla}^{\dot{\alpha}} \overline{{\cal C}} ]_D. 
\ee
Again,  by gauge fixing $S_0=1$ we go to the old-minimal supergravity, and the Lagrangian  (\ref{OM13}) is written as
\be
\nn
{\cal{L}} &=& \int {\rm d}^2 \Theta \, \, 2 \, {\cal{E}} \left\{
\frac{3}{8} \Big{(} \overline{{\cal{D}}}\overline{{\cal{D}}} - 8 {\cal{R}} \Big{)} e^{ - K/3 } + W 
\right\}+ {\rm h.c.}
\\
\label{OM14}
&+&\frac{\xi}{ \lambda_1^2} \int {\rm d}^2 \Theta \, \, 2 \, {\cal{E}} \left\{
\left(-\frac{1}{8}\right) \Big{(} \overline{{\cal{D}}}\overline{{\cal{D}}} - 8 {\cal{R}} \Big{)} 
{\cal{D}} ^{\alpha} {\cal C} {\cal{D}}_{\alpha} {\cal C} 
\overline{{\cal{D}}}_{\dot{\alpha}} \overline{{\cal C}} \ \overline{{\cal{D}}}^{\dot{\alpha}}  \overline{{\cal C}}
\right\}+ {\rm h.c.}, 
\ee
with K\"ahler potential
\be
\label{K6}
K = -3\, \text{ln}\,\big{(}{\cal T} + \overline{{\cal T}} - {\cal C} \overline{{\cal C}}\big{)} 
\ee
and superpotential
\be
\label{W6}
W=  \frac{3}{\sqrt{\lambda_1}} {\cal C} \left({\cal T} - \frac{1}{2}\right).
\ee
Theories of the form (\ref{OM14}) have been discussed in Ref. \cite{Cecotti:1986jy} 
and more recently extensivelly studied in \cite{Farakos:2013bca,EP}.
After integrating out the auxiliary fields (except $F_c$, the auxiliary field of the ${\cal C}$ superfield), 
and performing the rescalings, the Lagrangian  becomes
\be
\label{OM15}
e^{-1}{\cal L} &=& \frac{1}{2}  R  - K_{i {\overline{j}}} \partial_{\mu} z^i \partial^{\mu} \overline{z}^{\overline{j}} 
+\frac{16 \xi}{\lambda_1^2} \partial_{\mu} C \partial^{\mu} C \partial_{\nu} \overline{C} \partial^{\nu} \overline{C} 
- V_T +  {\cal L}_{F_c}, 
\ee
with
\be
\label{LFC}
{\cal L}_{F_c} &=& {\cal A} F_c + \overline{{\cal A}} \  \overline{F}_c
+ {\cal B}  F_c \overline{F}_c + {\cal S}  (F_c \overline{F}_c)^2,
\\
\label{VT}
V_T &=& e^{K} \frac{1}{K_{T\overline{T}}} D_T W D_{\overline{T}}\overline{W} - 3 e^{K} W \overline{W}
\ee
and
\be
{\cal A} &=& e^{2K/3} \frac{K_{C {\overline{T}}}}{K_{T \overline{T}}} D_T W - e^{2K/3} D_C W,
\\
{\cal B} &=& e^{K/3} K_{C {\overline{C}}} 
- e^{K/3} \frac{K_{T {\overline{C}}}K_{C {\overline{T}}}}{K_{T {\overline{T}}}}
-\frac{32 \xi}{\lambda_1^2}  e^{K/3} \partial^{\mu} C \partial_{\mu} \overline{C},
\\
{\cal S} &=& \frac{16 \xi}{\lambda_1^2} e^{2K/3}.
\ee
The equations of motion for $F_c$ are 
\be
0={\cal A} + {\cal B} \overline{F}_c + 2 {\cal S}   F_c \overline{F}_c^2 \ \ , \ \ 
0=\overline{{\cal A}} + {\cal B} F_c + 2 {\cal S}   \overline{F}_c F_c^2,
\ee
which can be combined into the single equation  
\be
\alpha = X (1 + \beta X )^2,
\ee
where
\be
\alpha &=& \frac{{\cal A} \overline{{\cal A}}}{{\cal B}^2},
\\
\beta &=& \frac{2 {\cal S}}{{\cal B}},
\\
X &=& F_c \overline{F}_c.
\ee
The solution to the above equation is then easily found to be
\be
X = \frac{2}{3 \beta} \big{(}\cosh m -1\big{)},
\ee
with
\be
m = \frac{1}{3 } \arccosh\Big{(}\frac{27}{2} \alpha \beta +1\Big{)}.
\ee
The final scalar potential will have the following compact form
\be
V_F = {\cal B} X + 3 {\cal S} X^2 + V_T.
\ee
To study the implications of the corrections on the inflaton potential we look again at the minimum 
$C=\bar{C}=b=0$ with the redefinition (\ref{T}). 
The inflaton field $\phi$
will now have a potential
\be
\label{VF9}
V_F = \frac{3 e^{-4 \phi/\sqrt{3}}}{8 s} \ \text{cosh}\ \frac{u}{3} \  \left( \text{sinh}\ \frac{u}{6}\right)^2,
\ee
with
\be
u = \arccosh\left\{ 1 + 36 \Big{(}-1 + e^{2 \phi/\sqrt{3}}\Big{)}^2 s  \right\}, \ \ s=\frac{\xi}{\lambda_1^3}.
\ee
The potential (\ref{VF9}) has been plotted  in Fig. 2 for various values of the parameter
$s$. If $s=0$ (i.e, $\xi=0$), the potential has a plateau for large positive values 
of $\phi$. For non-zero values of $s$, the plateau is restricted to smaller regions of the scalar field and, 
like for the new-minimal version,  it disappears for larger values of $s$ with a fall-off 
$V_F \sim e^{-4 \phi / \sqrt{3}}  $ after the plateau. 
\begin{figure}[H]
\begin{center}
\includegraphics[scale=0.6]{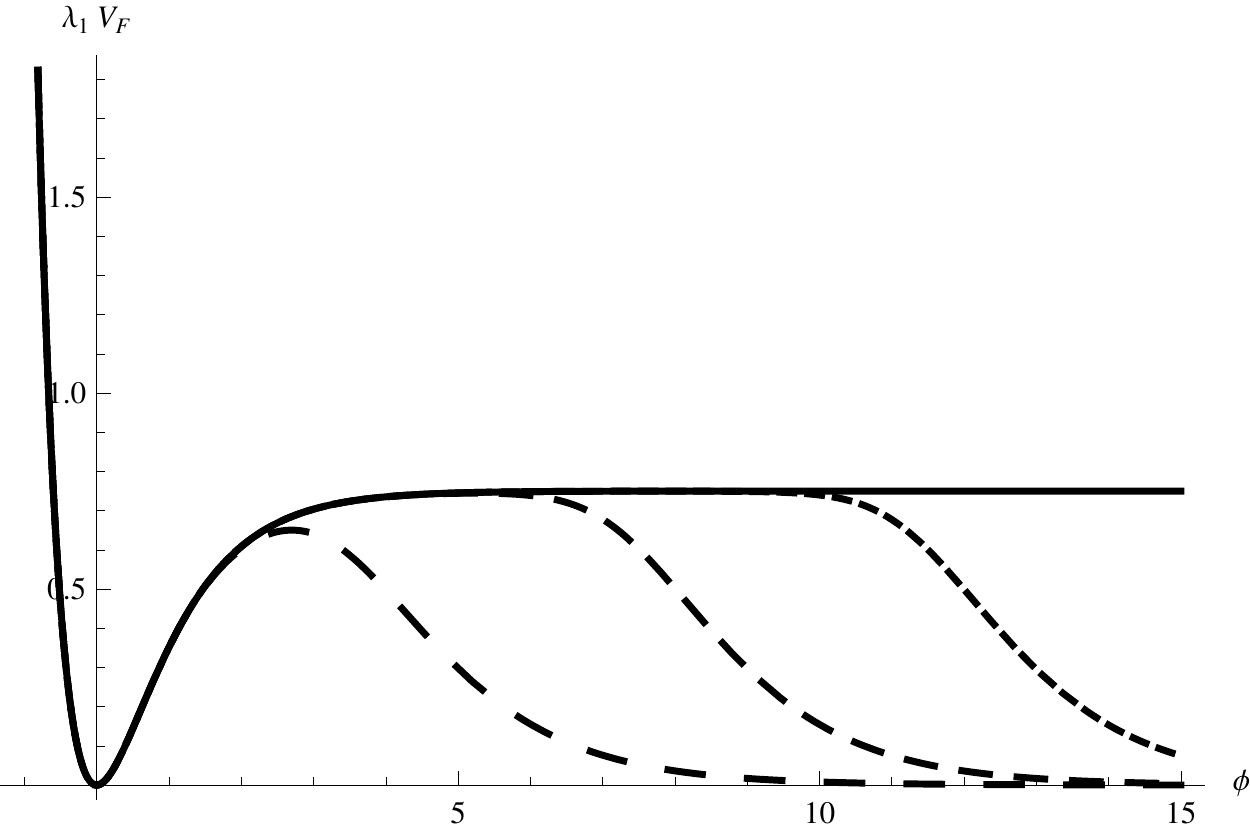}
\caption{\small The scalar potential   for three different values of $s=\frac{\xi}{\lambda_1^3} $: \, 
$i$) $s= 10^{-4}$ (long dashed line), $ii$) $s=10^{-8}$ (medium dashed line) and $iii$)
$s=10^{-12}$ (short dashed line). The horizontal line corresponds to $\xi=0$.}
\end{center}
\end{figure}
\noindent

\section{Conclusions}
In this paper we have discussed the embedding of the Starobinsky model of inflation within
${\cal{N}}=1$ supergravity. We have shown that the Starobinsky model can be derived from 
the   new-minimal supergravity, where a linear compensator superfield 
is employed. The  
Starobinsky model  becomes equivalent to standard supergravity coupled to a massive vector multiplet whose  
lowest scalar component plays the role of the inflaton and the vacuum energy is provided by  a  
 $D$-term potential.  We have subsequently investigated the robustness of the model against higher-order corrections allowed by the symmetries and concluded that they may represent a threat to the success of the model as they may destroy the flatness of the potential. This is true both in the old- and in the new-minimal formulation. In this sense, the Starobinsky model suffers from the 
same  difficulty  one encounters when trying to embed a model of inflation in supersymmetry  where the  flatness of the potential is easily destroyed by supergravity corrections \cite{lr}.

\section*{Aknowledgements}

The authors thank U.~Lindstr\"om for discussion. 
 A.R. is supported by the Swiss National
Science Foundation (SNSF), project ``The non-Gaussian Universe" (project number: 200021140236).
This research was implemented under the “ARISTEIA” Action of the “Operational Programme 
Education and Lifelong Learning” and is co-funded by the European Social Fund
(ESF) and National Resources. 
This work is partially supported by European Union's Seventh
Framework Programme (FP7/2007-2013) under REA grant agreement n°329083.

\end{document}